\documentclass[]{emulateapj} 

\usepackage{color}
\usepackage{subfigure}

\newcommand{\be}{\begin{equation}}
\newcommand{\ee}{\end{equation}}
\newcommand{\ber}{\begin{eqnarray}}
\newcommand{\eer}{\end{eqnarray}}
\newcommand{\bers}{\begin{eqnarray*}}
\newcommand{\eers}{\end{eqnarray*}}

\begin{document}

\title{Gamma-ray Bursts Induced by Turbulent Reconnection}

\author{Alex Lazarian}
\affil{Department of Astronomy, University of Wisconsin, 475 North Charter Street, Madison, WI 53706, USA; 
lazarian@astro.wisc.edu}

\author{Bing Zhang}
\affil{Department of Physics and Astronomy, University of Nevada Las Vegas, NV 89154, USA; zhang@physics.unlv.edu}
\affil{Department of Astronomy, School of Physics, Peking University, Beijing 100871, China}
\affil{Kavli Institute for Astronomy and Astrophysics, Peking University, Beijing 100871, China}

\author{Siyao Xu}
\affil{Department of Astronomy, University of Wisconsin, 475 North Charter Street, Madison, WI 53706, USA; 
Hubble Fellow;  sxu93@wisc.edu }

\begin{abstract}

We consider a simple model for gamma-ray bursts induced by magnetic reconnection in turbulent media. The
magnetic field in a jet is subject to kink instabilities, 
which distort the regular structure of the spiral magnetic field, drive turbulence, and trigger reconnection. 
The resulting reconnection takes place in a high Reynolds number medium, 
where turbulence is further enhanced and in turn accelerates the reconnection process. 
This boot-strap reconnection gives rise to bursts of reconnection events, through which 
the free energy of magnetic field is transformed into a gamma-ray burst.
The efficiency of magnetic reconnection and magnetic energy dissipation is not constrained by microphysical properties of plasmas.

\end{abstract}


\section{Introduction}

Gamma-ray bursts (GRBs) are the most energetic phenomena in the modern Universe.
The physical mechanism to produce the observed $\gamma$-ray emission is still not
identified (e.g. Kumar \& Zhang 2015 for a recent review). Here we consider a scenario in which the magnetic reconnection
in turbulent media induces GRBs. We employ Lazarian \& Vishniac (1999,
henceforth LV99) model of reconnection which has recently been extended for the relativistic regime
(Takamoto et al. 2015, Lazarian et al. 2016).  
This model proposes that the reconnection rate depends on the intensity of the surrounding turbulence,
and the reconnection is a boot-strap process as the reconnection-driven turbulence acts to boost the reconnection efficiency.

In the standard fireball model (Pac\'ynski 1986; Goodman 1986; Shemi et al. 1990; Rees \& M\'esz\'aros 1992, 1994;
M\'esz\'aros \& Rees 1993, 1997, 2000), magnetic fields 
are not dynamically important, i.e. $\sigma \ll 1$ in the emission region, where $\sigma$ is the ratio between the 
Poynting flux and the matter (baryonic $+$ leptonic) flux.  
As an alternative picture that is getting more and more popular,
the magnetic field is dynamically important
in GRB outflows, i.e. $\sigma_0 \gg 1$ at the central engine, and $\sigma \geq 1$ in the emission
region (see, e.g., Usov 1992;  Thompson 1994; Lazarian et al. 2003; Lyutikov \& Blandford 2003;
Zhang \& Yan 2011, henceforth ZY11). In this model, the GRB emission is powered by the magnetic energy
dissipation within the ejecta. 
Evidence supporting a Poynting-flux-dominated outflow in at least some GRB jets includes: the lack of an 
observed weak thermal component in most GRB spectra (Zhang \& Pe'er 2009); strongly polarized GRB
emission (Coburn \& Boggs 2003; Willis et al. 2005; Yonetoku et al. 2011, 2012) and early optical emission
(Steele et al. 2009; Mundell et al. 2013; Troja et al. 2017); more and more stringent upper limit of high-energy 
neutrino emission from GRBs (Zhang \& Kumar 2013; Aartsen et al. 2015, 2016, 2017); 
and evidence of bulk acceleration or anisotropic emission in GRB prompt emission and X-ray flares (Uhm \& Zhang 2016a,b;
Jia et al. 2016; Geng et al. 2017a). A natural mechanism to dissipate magnetic energy is through magnetic reconnection.

Magnetic reconnection has been widely discussed as the energy dissipation mechanism for 
GRBs (see Lyutikov \& Lazarian 2013 for review and ref. therein).  
However, 
the problem lies in the intrinsic difficulty of
reconnection since it tends to be a very slow process in ordered magnetic fields. 
In addition, as in the case
of solar flares, both a slow phase and a fast bursty phase of reconnection are required for explaining GRBs. 
Turbulent reconnection was suggested to account for the GRB observations in 
Lazarian et al. (2003), where the LV99 model was employed. 
It was conjectured that LV99 can be generalized for the relativistic case 
and the reconnection can proceed with $V_A\rightarrow c$.
More recent research provided support for this conjecture (Takamoto et al. 2015).

As discussed in Lazarian et al. (2003), 
along with the build up of turbulence, 
reconnection rate increases. 
It induces a positive feedback and further drives turbulence, 
resulting in the explosive reconnection. 
This idea became the basis of the Internal-Collision-induced MAgnetic Reconnection 
and Turbulence (ICMART) model (ZY11), who showed that such a model can overcome several difficulties
of the traditional internal shock model (Rees \& M\'esz\'aros 1994; Kobayashi et al. 1997; Daigne \& Mochkovitch 1998;
Ghisellini et al. 2000; Kumar \& McMahon 2008) and can well interpret the lightcurves and spectra of GRBs 
(Zhang \& Zhang 2014; Uhm \& Zhang 2014; Xu \& Zhang 2017; Xu et al. 2017).

ZY11 speculated that the magnetic field reversals required to trigger ICMART events may be achieved through
internal collisions among high-$\sigma$ blobs. Under the framework of a helical magnetic configuration, 
they suggested that repeated collisions may accumulate magnetic distortions and eventually reach the 
threshold to trigger the run-away turbulent reconnection. Deng et al. (2015) performed a series of relativistic MHD
numerical simulations of collisions of high-$\sigma$ magnetic blobs, and found that significant magnetic dissipation
can indeed occur with an efficiency above 30\%. However, the simulations are on the global scale and no detailed
turbulent reconnection features on small scales can be observed.

In this paper we introduce an alternative mechanism to trigger ICMART-like events in ZY11 by invoking the kink instability. 
This can provide a self-consistent scenario of GRBs based on the turbulent reconnection model. 
As the main difference of this model
from other kink-driven models of GRBs (e.g. Drenkhahn \& Spruit, 2002; Giannios \& Spruit, 2006; Giannios, 2008; McKinney \& Uzdensky, 2012), 
the kink instability also induces turbulence (Galsgaard \& Nordlund, 1997; Gerrard and Hood, 2003), which drives fast magnetic reconnection 
similar to the original model of ZY11. 
Regarding the theoretical basis of our model, 
we will discuss the details of turbulent reconnection in view of the latest theoretical and numerical advances in the high-$\sigma$, relativistic regime.

In what follows we first present the physical ingredients of our model in \S 2 and 
justify the applicability of turbulent reconnection process in GRB conditions in \S 3. 
We confront our model with GRB observations in \S 4. 
A discussion of our results is provided in \S 5.

\section{Model Ingredients}
\subsection{Triggering magnetic dissipation through kink Instability}\label{ssec: kik}

Various theoretical arguments and observational evidence suggest that GRBs originate from ultra-relativistic jets with bulk Lorentz factor $\Gamma > 100$ (e.g. Lithwick \& Sari 2001; Taylor et al. 2004; Zhang et al. 2006; Kato et al. 2008; Abdo et al. 2009a,b). 
Various polarization studies of prompt and afterglow emission indicate the presence of a large-scale ordered magnetic field (Yonetoku et al. 2011, 2012; Mundell et al. 2013; Wiersema et al. 2014) in the ejecta. Rotation is a generic property of astrophysical jets that arises from the transfer of the angular momentum from the accreting material and the central engine through the magnetic field (see Blandford \& Znajek 1977, Blandford \& Payne 1982; Bisnovatyi-Kogan \& Lovelace 2001). 
Such a rotation is expected to produce a magnetic spiral within the jet associated with a GRB (see more in Kumar \& Zhang 2015).  
Although the magnetic spiral has substantial free energy, a helical magnetic field does not reconnect on its own.
Indeed, magnetic fields in the adjacent magnetic coils are of the same direction, 
and this inhibits their reconnection.  
Therefore, the spiral should be destabilized to allow for the magnetic reconnection.

A kink instability is one of the plausible processes that can destabilize the spiral magnetic field in the jet. 
The stability of current-carrying force-free (or nearly force-free) fields was extensively studied for cylindrical geometry in the astrophysical context (see Baty \& Heyvaerts 1996; Li 2000; Baty 2001; Gerrard et al. 2002, Torok et al. 2004). 
Physically, the kink instability arises as the winding of the magnetic field in the jet gets so tight that 
a particular threshold is exceeded. 
This process can be induced if, for instance, the jet is slowed down by the external media, 
or there are variations of velocity and density within the jet. 
As this happens, the adjacent coils of magnetic fields get closer and subsequently the kink instability develops. 
It results in oppositely directed magnetic fluxes that can reconnect releasing the stored magnetic energy.

\begin{figure}[htbp]
\centering
   \includegraphics[width=8cm]{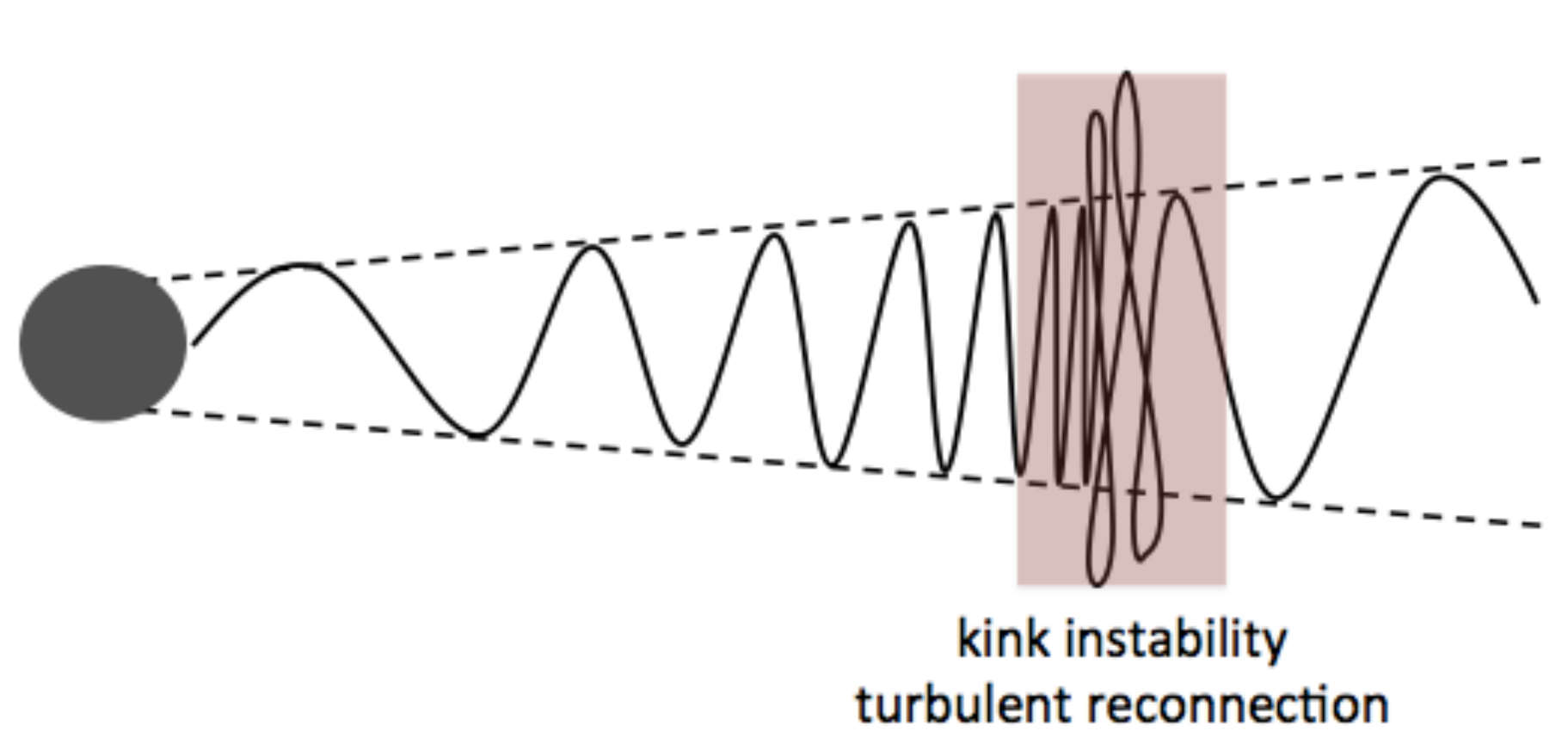}
\caption{Illustration of the kink instability in the GRB jet.}
\label{figure2}
\end{figure}

Figure \ref{figure2} illustrates the kink instability of the magnetic field in the GRB jet. In this scenario,  
due to the velocity variations within the strongly magnetized jet launched by the central engine,
the faster part of the jet approaches its slower part in front, where the spiral magnetic field is squeezed together 
and the condition for triggering the kink instability (see Eq. (\ref{eq:2})) is satisfied. 
The resulting magnetic flux reversals entail magnetic reconnection. In the presence of turbulence,
it is substantially efficient in energy dissipation to account for the GRB emission 
(see \S \ref{ssec: returr}).

For simplicity, let us consider a cylindrical jet with a length $L$ and a cross-section radius $R$. 
In the case of a helical magnetic field geometry, the spiral magnetic field obeys the equation (see Freiberg 1987):
\begin{equation}
\frac{R d\theta}{dz}=\frac{B_t}{B_p},
\end{equation}
where $B_p$ and $B_t$ are the poloidal and toroidal magnetic field strengths, respectively.
$\theta$ gives the toroidal direction, and $z$ is the distance along the jet axis. 
One can then define a safety factor $q$, such that $2 \pi R / q L = B_t/B_p$ is satisfied. The kink instability condition is given by
\begin{equation}
q = \frac{2\pi R B_p}{L B_t} < 1,
\label{1}
\end{equation}
which is called the Kruskal-Shafranov (KS) criterion. This can be rewritten as
\begin{equation}
 \frac{ B_t}{B_p}  > \frac{2\pi R}{L}.
\label{eq:1}
\end{equation}
The growth rate of the instability
is $\gamma_g \sim \frac{B_0}{\rho R} \sqrt{(1/q) (1/q-1)}$
(see more in Goedbloed \& Poedts 2004). 
It grows 
faster for a stronger magnetic field $B_0  \sim B_t$ (for toroidally-dominated field) and 
a lower plasma density $\rho$.

The above idealized criterion only applies to the situation with a constant $\rho$ and uniform winding of magnetic fields. 
It should be modified in realistic settings of GRBs with more complex structure of density and magnetic fields. 
Besides, in the relativistic case, 
it is appropriate to adopt the force-free approximation where only the charges, currents, and fields are accounted for, 
but the inertia and pressure of the plasma are ignored. 
By taking into account the stabilizing effect of the rotating magnetic spiral,
one can extend the classical KS criterion.
Under this consideration, we find that the kink instability arises if both the KS criterion and an additional condition
(Tomimatsu et al. 2001)  
\begin{equation}
 \frac{B_t}{B_p} > \frac{R}{R_{\rm LC}} = \frac{R \Omega_B}{c}
 \label{eq:2} 
 \end{equation}
are satisfied. 
Here $\Omega_B$ is the angular velocity of the magnetosphere of the central engine, 
which is also the angular velocity of the spiral magnetic field, 
and $R_{\rm LC} = c/\Omega_B$ is the radius of the light cylinder of the central engine.
For GRBs, $L = c \Delta t_{\rm slow} = 3\times 10^{10} \ {\rm cm} (\Delta t_{\rm slow})$, 
where $\Delta t_{\rm slow}$ is the typical duration of the ``slow variability component'' of GRB lightcurves, which defines the duration of central engine activity for each active episode of GRB emission (Gao et al. 2012). The central engine of GRBs are typically millisecond rotators, so that $R_{\rm LC} = 4.8 \times 10^6 {\rm cm} P_{-3}$. As a result, the criterion Eq.(\ref{eq:2}) is more stringent than Eq.(\ref{eq:1}), 
so that it is more relevant.

More complicated set-ups for the kink instability development within relativistic jets have been explored numerically (see McKinney \& Blandford 2009, Mizuno et al. 2009, 2011, 2012, 2014, Mignone et al. 2010, ONeill et al. 2012). These simulations revealed a variety of initial conditions that influence the growth and the evolution of the kink instability. The numerical results are consistent with the general conditions 
Eqs. (\ref{eq:1}) and (\ref{eq:2}), which we use as a guidance for our further discussion. 

For a Kerr black hole as the central engine that launches a relativistic jet, the jet is kink stable if the condition 
\begin{equation}
 | \Omega_B - \Omega_{\rm BH} | < \Omega_{B}
\end{equation}
is satisfied (Tomimatsu et al. 2001). This requires that the magnetosphere angular velocity does not differ from the black hole angular velocity significantly, that is (Tomimatsu et al. 2001)
\begin{equation}
 \Omega_B \geq \Omega_{\rm BH}/2.
\end{equation}
Such a condition is usually satisfied for a GRB engine, so that for a helical jet launched from a hyper-accreting BH, the jet may propagate to a large distance without triggering kink instability.

For a steady cylindrical jet, from Eq.(\ref{eq:2}) one can see that the kink instability condition is either satisfied or not throughout the jet propagation. Such a conclusion applies even in a more general case with the jet radius evolving with the distance from the central engine. 
Without losing generality, one can write
\begin{equation}
 R \propto r^b.
\end{equation}
For a cylindrical and a conical jet, one has $b=0$ and $b=1$, respectively. In general, one may have $0<b<1$. 
Magnetic flux conservation gives 
\begin{eqnarray}
B_p & \propto &  R^{-2} \propto r^{-2b}, \\
B_t & \propto  & R^{-1} \propto r^{-b}.
\end{eqnarray}
This suggests that both sides in Eq.(\ref{eq:2}) are proportional to $r^b$, so that Eq.(\ref{eq:2}) is satisfied (or not) throughout the jet regardless of the geometrical configuration of the jet. 
It shows that in order to trigger kink instability in a jet, one needs to introduce additional mechanism to alter the magnetic configuration of the jet.

There are at least three possible ways of triggering kink instability in the GRB context. 
In the first scenario, a magnetized jet is decelerated as it penetrates through the stellar envelope of the progenitor star. 
This would induce significant magnetic energy dissipation below the photosphere and result in a matter-dominated fireball 
with strong photospheric emission. 
It more likely happens during the early phase of a GRB. 
At later times when the early portion of the jet successfully escapes the star, 
the Poynting-flux-dominated jet is able to reach a large distance from the central engine before significant dissipation happens. 
The second scenario to trigger kink instability involes external pressure from the ambient medium. 
Analogous to the external shock model of GRBs (Rees \& M\'esz\'aros 1992; M\'esz\'aros \& Rees 1993, 1997), it invokes external medium to 
decelerate the jet, and thus, the increase of the $B_t/B_p$ ratio in the jet triggers kink instability. 
The emission region of this model is close to the deceleration radius, i.e. $R_{\rm GRB} \sim R_{\rm dec} \sim 10^{17}$ cm for typical GRB parameters. 
The third scenario, similar to the internal shock model (Rees \& M\'esz\'aros 1994) and the ICMART model (ZY11) of GRBs, 
requires intrinsic irregularity of the central engine and interactions between different parts in the jet with different bulk Lorentz factors to increase the $B_t/B_p$ ratio to trigger kink instability (see Figure \ref{figure2}). 
This latter scenario is more consistent with the GRB observational data (see \S\ref{sec:obs} for more discussions). 
Similar to the ICMART model (ZY11), the emission radius in this scenario is $R_{\rm GRB} \sim \Gamma^2 c \Delta t_{\rm slow} \sim 10^{15}$ cm. 
This can be understood based on the following reasons. 
In the ICMART model, the collision of two magnetized shells is responsible for altering the magnetic configuration and triggering reconnection. 
In the current scenario, instead of the collision of two physically separated magnetized shells, 
it simply requires a continuous jet with velocity fluctuations within it. 
The trailing high-$\Gamma$ part of the jet catches up with the leading low-$\Gamma$ part at a radius similar to the collision one. 
Without a direct collision, 
the ram pressure of the trailing part squeezes the magnetic field configuration in the system, 
leading to the onset of kink instability and magnetic dissipation. 
A GRB is then produced around the same radius as the ICMART model.

It is important to note that kink instability does not necessarily disrupt the jet, but only results in the change of magnetic field structure, 
which enables the subsequent magnetic reconnection. 
Magnetic reconnection is driven by the free energy of magnetic fields. 
In a generic situation of 3D geometry, it causes the annihilation of 
contacting oppositely directed magnetic fluxes.

\subsection{Relativistic Reconnection of Turbulent Magnetic Fields}
\label{ssec: returr}

\subsubsection{Magnetic reconnection in turbulence}

\begin{figure*}[htbp]
\centering
   \includegraphics[width=10cm]{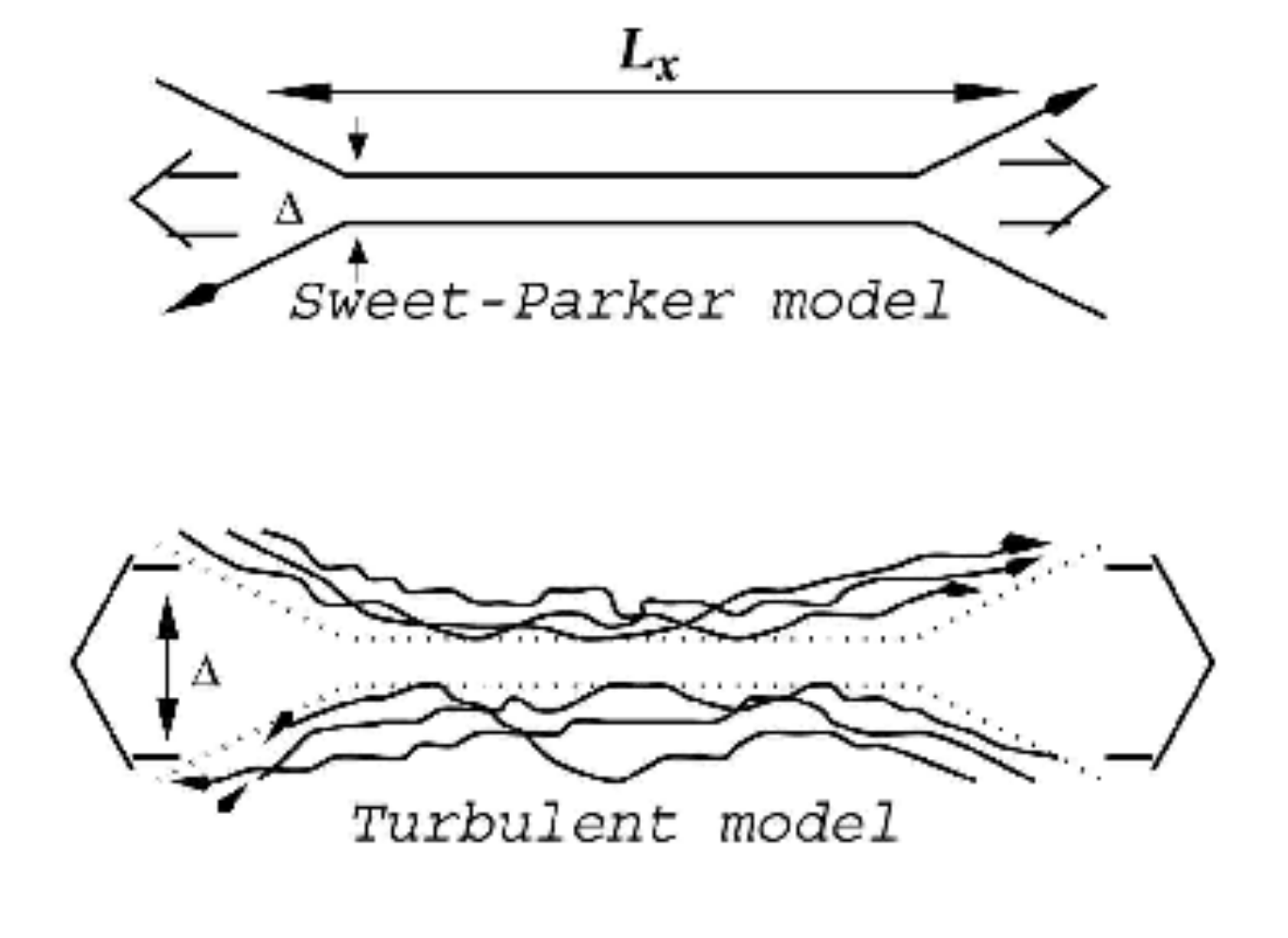}
\caption{Upper panel: Sweet-Parker reconnection. Lower Panel: Turbulent reconnection in LV99. }
\label{figure1}
\end{figure*}

The problem that challenges the traditional reconnection model, known as the Sweet-Parker model (shown in the upper part of Figure \ref{figure1}), 
is the unrealistically slow reconnection rate in astrophysical conditions.
This inefficiency arises from the disparity between the astrophysical scale $L_x$, 
over which the plasma is carried into the reconnection region,
and the microphysical scale $\Delta$ determined by the plasma resistivity, 
over which the plasma is ejected out from the reconnection region. 
Taking into account that the ejection velocity is approximately the Alfven velocity $V_A$, 
one can easily find that the reconnection rate for incompressible media, 
\begin{equation}
V_{rec}\approx  V_A \frac{\Delta}{L_x},
\label{for1}
\end{equation}
is very small, $\ll V_A$. 
In fact, for the outflow region determined by the Ohmic resistivity $\Delta \approx \eta/V_{rec}$, one recovers the
Sweet-Parker formula for the reconnection rate $V_{rec, SP}\approx V_A S^{-1/2}$.
Here $S= L_xV_A/\eta$ is the Lundquist number, where $\eta$ is the resistivity.
It can be huge, e.g. of the order $10^{10}$ or even $10^{20}$, in many astrophysical situations. 
As a result, the reconnection rate in the classical Sweet-Parker model is negligible for typical astrophysical settings.  

Below we show that the situation changes dramatically in the presence of turbulence. 
Turbulence is ubiquitous in astrophysical environments, 
and it is detected essentially in every case where it is searched for,
e.g., the Big Power Law in the Sky of interstellar electron density fluctuations 
(Armstrong et al. 1993, Chepurnov \& Lazarian 2010),  
non-thermal line-width broadening of various spectral lines. 
As we will describe later, 
there are strong reasons for us to expect that the reconnection in GRB environments takes place in a turbulent medium.

Turbulence is stochastic, but it obeys statistical laws. 
The famous Kolmogorov scaling is an example of such a law. 
For incompressible MHD turbulence, 
an analog of Kolmogorov theory is the theory proposed in 
Goldreich-Sridhar (1995, henceforth GS95; see Brandenburg \& Lazarian 2013 for a review)
\footnote{We believe that the time of vigorous debates of whether the GS95 model should be modified, e.g. by taking into account additional effects like alignment/polarization (Boldyrev 2005, 2006, Beresnyak \& Lazarian 2006),
non-locality of turbulence (Gogoberidze 2007), is over. 
Both theoretical (Beresnyak \& Lazarian 2010) and numerical (Beresnyak 2013, 2014) studies suggest that the GS95 model provides 
a proper description of MHD turbulence. 
Therefore in what follows we do not discuss alternative turbulence models. 
In any case, the insignificant changes of the scalings, e.g. from the Kolmogorov spectrum of $k^{-5/3}$ to the Kraichnan spectrum of $k^{-3/2}$ 
advocated by alternative constructions, do not change significantly the model of turbulent reconnection in our consideration.}. 
The GS95 theory is the basis of 
Lazarian \& Vishniac (1999) theory of magnetic reconnection.

In what follows we shall employ the model of turbulent reconnection in LV99. 
This model of non-relativistic reconnection has been numerically tested in Kowal et al. (2009, 2012) 
and successfully compared with observations in a number of studies 
(see Ciravella \& Raymond 2008, Sych et al. 2009, Eyink et al. 2013, Eynk 2014, Singh et al. 2015, Kadowaki et al. 2015, Khali et al. 2015, Lalescu et al. 2015, see also Lazarian et al. 2015, 2016 for reviews)
\footnote{We note that in the review by Karimabadi \& Lazarian (2013), it was stated that no studies revealed the correspondence between the observed Solar wind reconnection and the LV99 predictions. 
This deficiency was corrected in Lalescu et al. (2015) where such correspondence was found.}. 
Note that the LV99 expression that we will apply has also been re-derived using other theoretical approaches 
in Eyink et al. (2011) and Eyink (2015). 
This model was employed in Zhang \& Yan (2011) and became the corner stone of the ICMART model for GRBs. 
More recent study in Takamoto, Inoue \& Lazarian(2015, henceforth TIL15) showed that the LV99 model can be successfully generalized for the relativistic reconnection. 

The LV99 model and its comparison with the classical Sweet-Parker reconnection is illustrated in Figure \ref{figure1}.  
Within LV99 model,  the outflow region is determined by the magnetic field line wandering. 
This is in contrast to the Sweet-Parker model where the outflow is determined by the plasma microscopic diffusivity. 
As a result, within the LV99 model the reconnection can be both fast and slow depending on the level of turbulence. 
If turbulence is of low amplitude, the magnetic field wandering is small, and thus the resulting outflow opening $\Delta$ is strongly constrained.
The reconnection speed can be obtained from the mass conservation, 
namely, $V_{rec}\approx V_A \Delta/L_x$, where $V_A$ is the Alfven speed, 
and $L_x$ is the extent of contact surface between two magnetic fluxes in Fig. \ref{figure1}. 
The outflow thickness grows as the level of turbulence increases. 
Naturally, this increases the rate of turbulent reconnection. 
The fact that the reconnection can be both slow and fast is important for explaining reconnection explosions 
that we claim is a part of the GRB phenomenon. 

Wandering or meandering of magnetic field is well known 
(see Jokipii 1973) and numerically tested (see Lazarian, Vishniac \& Cho 2004, Beresnyak 2011). 
This effect has been applied for decades to studying the perpendicular diffusion of cosmic rays in astrophysical magnetic fields, 
although the proper quantitative treatment of the effect was only developed in LV99.
LV99 not only serves as a theory for the turbulent reconnection, but also provides a 
different physical interpretation of GS95 theory of turbulence. 
In particular, one can describe turbulent motions as eddies perpendicular to the magnetic field. 
The induced mixing of field lines in the direction perpendicular to the magnetic field 
is most energetically favorable as it does not involve magnetic field bending. 
Such mixing motions of field lines are facilitated by the 
turbulent reconnection that acts within one eddy turnover time. 
While in the original GS95 paper, 
the mixing motions were believed to be in the direction perpendicular to the mean magnetic field, 
LV99 correctly pointed out that it should be local magnetic field that matters. 
Naturally, as the perpendicular mixing is not subject to the magnetic tension, it leads to the energy cascade consistent with the 
Kolmogorov spectrum.
As the other corner stone of GS95 theory, the scaling relation of anisotropic MHD turbulence can also be easily understood 
in LV99 picture. It is the consequence of the equalization between the period of the Alfvenic perturbation 
along the magnetic field and the eddy turnover time.

By extending the GS95 theory to the sub-Alfvenic regime, LV99 obtained the expression of $\Delta$ from magnetic field wandering:
\begin{equation}
\Delta\approx L_x \left(2\epsilon_{inj} l {V_A^3}\right)^{1/2} \min\left[\left({L_x\over l}\right)^{1/2},
\left({l\over L_x}\right)^{1/2}\right],
\label{Delta}
\end{equation}
where $\epsilon_{inj}$ is the injected turbulent energy,
and $l$ and $L_x$ are the turbulence injection scale and the extend of the ``current sheet".
We term it as ``current sheet", becuase in turbulent media the individual current sheet evolves to produce a complex network of fractal current sheets, 
which extends over the thickness $\sim \Delta$ determined by magnetic field line wandering.
The speed of reconnection $V_{rec}$ can trivially be obtained from the mass conservation condition:
\begin{equation}
\rho_i V_{rec} L_x =\rho_s V_A \Delta,
\label{mass}
\end{equation}
where $\rho_i$ is the density of the inflow and $\rho_s$ is the density of the matter in the ``current sheet". 
Under the incompressible approximation adopted in LV99, there is $\rho_s=\rho_i$, and thus
turbulent reconnection rate can be presented as
\begin{equation}
V_{rec}\approx V_A \left(2\epsilon_{inj} l{V_A^3}\right)^{1/2} \min\left[\left({L_x\over l}\right)^{1/2},
\left({l\over L_x}\right)^{1/2}\right].
\label{recon1}
\end{equation}

For sub-Alfvenic driving, the injection of energy is related to the velocity at the injection scale $V_l$ according to the expression (LV99)
\begin{equation}
\epsilon_{inj}\approx V_l^4/2 lV_A.
\label{eps}
\end{equation}
Combining Eqs. (\ref{recon1}) and (\ref{eps}), one can get 
\begin{equation}
V_{rec}\approx V_A\min\left[\left({L_x\over l}\right)^{1/2},
\left({l\over L_x}\right)^{1/2}\right] \left(\frac{V_l}{V_A}\right)^2,
\label{recon}
\end{equation}
which indicates that $V_{rec}$ of turbulent reconnection differs from $V_A$
by a factor that depends on the ratio between the turbulence injection scale to the current sheet scale, 
as well as the ratio between the velocity at the injection scale and the Alfven velocity.

\subsubsection{Magnetic reconnection in relativistic and strongly magnetized turbulence}

For GRBs, we deal with relativistic plasmas with strong magnetization $\sigma\equiv B^2/4\pi h c^2\gg 1$, 
where $h=4$ is the specific enthalpy of relativistic ideal gas.  
In what follows, we summarize our current knowledge about this regime of turbulent reconnection. 
 
Very importantly, 
the GS95 model can also be used for describing relativistic Alfvenic turbulence 
(Thompson \& Blaes 1998). 
This relativistic analog of GS95 was successfully tested for the case of decaying turbulence in Cho (2005) under the so-called force-free approximation.
\footnote{In relativistic MHD turbulence, the force free approximation corresponds to the zeroth term of expansion of relativistic magnetohydrodynamics over a small parameter $1/\sigma$, where $\sigma$ is the ratio $2u_B/u_\rho$ with $u_B=B^2/8\pi$ as the magnetic energy density and $u_\rho=\rho c^2$ as the rest mass energy density.} 
The simulations of fully relativistic MHD turbulence  (Zhang et al, 2009; Inoue et al, 2011; Beckwith and Stone, 2011; Zrake \& MacFadyen, 2012, 2013; Garrison \& Nguyen, 2015) delivered results also consistent with the GS95 expectations. 
Recent studies of compressible relativistic MHD turbulence in Takamoto \& Lazarian (2016, 2017) revealed the difference between it and its non-relativistic analog (see Cho \& Lazarian 2002, 2003, Kowal \& Lazarian 2010). 
While the scaling of the Alfven and slow modes are similar to those in the non-relativistic simulations, 
the coupling of the Alfven and fast modes is stronger in the relativistic case. 
This coupling requires further studies, and it presents an uncertainty factor in this work. 
However, it does not change the general correspondence between relativistic and non-relativistic MHD. 

In addition, the analogy between turbulence in the non-relativistic and relativistic regimes 
extends to a regime of imbalanced MHD turbulence with different energy fluxes in opposite directions. 
The corresponding theory of non-relativistic imbalanced or non-zero cross-helicity MHD turbulence was earlier suggested 
in Beresnyak \& Lazarian (2008). 
Furthermore, it was shown to be also true for imbalanced relativistic MHD turbulence 
(Cho \& Lazarian 2014). 
This supports our claim of the intrinsic similarity between Alfvenic turbulence in relativistic and non-relativistic regimes.

Based on the similarity between relativistic and non-relativistic Alfvenic turbulence, 
one can expect that magnetic reconnection also gets fast in relativistic magnetized fluids. 
Despite the difference,  
i.e., higher coupling strength between Alfvenic and fast modes as discussed earlier, physics of turbulent reconnection stays the same. 
Besides, 
magnetic field lines in relativistic case can also be traced by the charged particle trajectories, 
and the effect of increased outflow thickness can also remove the 
bottleneck of the Sweet-Parker reconnection in the relativistic situation.

Indeed, recent relativistic simulations in TIL15 have confirmed the similarity between the relativistic reconnection and the non-relativistic one. 
They demonstrated that the turbulent reconnection speed can be as large as $0.3 c$,
which thus enables highly efficient 
conversion of magnetic energy into kinetic motions and particle acceleration. 
The numerical results in TIL15 can be approximated by an expression that generalizes Eq. (\ref{recon}). 
To derive this expression, 
one should take into account both the density change in the relativistic plasma and the modification of turbulence properties in the relativistic regime. 
The former can be obtained from the conservation of energy flux.
With both analytical considerations and numerical simulations provided in TIL15, there is 
\begin{equation}
\frac{\rho_s}{\rho_i}\sim 1-\beta \left(\frac{V_l}{c_A}\right)^2,
\end{equation}
where $\beta$ was found in numerical simulations to be a function of $\sigma$. 
The change of $\Delta$ was shown to correspond to the original Eq. (\ref{Delta}),
but with $\epsilon_{inj}$ reduced compared to the value in non-relativistic case. 
This corresponds to the transfer of larger fraction of energy to the fast modes which are subdominant in inducing magnetic field wandering, as indicated by simulations in Takamoto \& Lazarian (2016, 2017). 
With these modifications, the corresponding expression of the turbulent reconnection can be written as
\begin{eqnarray}
V_{rec, relativ.} & \approx & V_A \left(\frac{\rho_s}{\rho_i}\right) \left(2\alpha \epsilon_{inj} l {V_A^3}\right)^{1/2} \nonumber \\
& \times & \min\left[\left({L_x\over l}\right)^{1/2},
\left({l\over L_x}\right)^{1/2}\right],
\label{relativ}
\end{eqnarray}
where $\alpha<1$ is the factor accounting for the decrease in the fraction of magnetic energy that induces magnetic field wandering. 

It is evident from Eq. (\ref{relativ}) that the theory of relativistic turbulent reconnection does require further development in order to decrease the uncertainties related to magnetic turbulence in relativistic fluids. 
For the time being it is important for our further discussion that qualitatively relativistic turbulent reconnection is similar to its non-relativistic counterpart. 
The existing numerical simulations in TIL15 provide us with guidance for studying the reconnection in GRBs. 
In particular, it is clear from the simulations that in high $\sigma$ flows, 
the turbulent reconnection speed only slowly changes with the injection velocity and does not depend on the guide magnetic field,
i.e., the common field component shared by the two reconnected fluxes.
The injection scale of the turbulence induced through the kink instability is likely to be comparable to the scale of the magnetic field flux tubes, 
i.e. $l\sim L_x$. 
In this situation, by extrapolating the results in TIL15, one can claim that the reconnection rate is larger than $0.1 c_A$, 
where the relativistic Alfven speed $c_A$ is very close to the light speed.

\section{GRBs driven by turbulent reconnection}

\subsection{Justification of turbulent reconnection in GRBs}

There are other theories proposed for increasing the turbulent reconnection rate. 
Therefore, it is necessary to discuss 
why we believe that the turbulent reconnection is the most relevant process 
for the GRB physics. 
 
The model suggested by 
Petscheck (1968) 
was for decades the well accepted mechanism for the fast magnetic reconnection. 
The mechanism reached the apogee of popularity
after numerical simulations including the Hall effect
show that the Petschek-type X-point reconnection can happen in a collisionless plasma 
(Shay et al, 1998; Drake, 2001; Drake et al, 2006). 
However, as pointed out in e.g. LV99,
such configurations are very difficult to realize in realistic astrophysical settings.

Another model for fast reconnection relies on the instabilities of the Sweet-Parker reconnection layer, e.g. tearing instabilities. 
Their importance was strongly advocated by Syrovatskii (see 1981 for a review and ref. therein) 
and has been widely recognized by the community more recently 
(Biskamp 1986; Shibata \& Tanuma 2001, Daughton et al. 2006, 2009, 2011, 2014; Fermo et al. 2012, Loureiro et al. 2007, 2012, Lapenta 2008; Bhattacharjee et al. 2009, Cassak et al. 2009; Huang \& Bhattacharjee 2010, 2012, 2013; Shepherd \& Cassak 2010; Uzdensky et al, 2010, Huang et al. 2011, Barta et al. 2011, Huang et al. 2011, Shen et al. 2011,  Takamoto 2013, Wyper \& Pontin 2014). 
The reconnection rates obtained in MHD regime were limited to $\sim 0.01$ of $V_A$
(e.g. Loureiro et al. 2007), 
which is obviously inadequate to account for the energy dissipation in GRBs.

Differently, we believe that turbulence plays a significant role in the context of the GRBs. 
The typical Reynolds number in GRB conditions is $Re\sim (10^{27} - 10^{28}) \gg 1$ (ZY11), 
where turbulence is inevitable and its effect on reconnection cannot be disregarded. 
For instance, it is well known that turbulence can efficiently suppress tearing instabilities (Somov \& Vornota 1993).
This was also numerically confirmed by Kowal et al. (2018).
Moreover, as we discussed earlier, 
turbulence enables fast relativistic reconnection without the need of any instabilities.

One can provide arguments that 
even initially the level of turbulence is low and tearing instabilities are more important, 
the generic final picture of reconnection will be dominated by turbulence. 
The relevant fast reconnection in astrophysics is that at a very large Lundquist number, i.e. $S\gg 1$. 
This number is related to the Reynolds number of the outflow  
$Re=\Delta V_A/\kappa$, where $\kappa$ is viscosity,  by 
\begin{equation}
Re=S \frac{V_{rec}}{V_A} Pt^{-1} ,
\label{Re}
\end{equation}
where $Pt=\kappa/\nu$ is the Prandtl number. 
Thus for $V_{rec}$ being $0.01$ or a larger fraction of $V_A$, 
the $Re$ number of the outflow increases in parallel with $S$. 
In GRB magnetic dissipation region with $\sigma > 1$, $S$ is essentially the magnetic Reynold's number in the Bohm diffusion limit, which is $\sim 3\times 10^{12}$ for typical GRB parameters (Eq.(40) of ZY11). 
Therefore, it is natural to expect the outflow to be turbulent and the transfer to turbulent reconnection to occur (see Lapenta \& Lazarian 2012).

Therefore, turbulent reconnection is the most likely
realization of the magnetic reconnection in the conditions of a GRB.  
Even current low-resolution numerical simulations show the development of turbulence as an outcome of 3D reconnection both in compressible and incompressible media (see Oischi et al. 2015, Lazarian et al. 2015, 2016, Beresnyak 2017). 
The transfer to the state of turbulence within 3D reconnection was also observed in Huang \& Bhattachargee 2016. 
A more extended study of the same set up in Kowal et al. (2017) showed that the GS95 turbulence is generated as a result of reconnection and the LV99-type reconnection ensues. 
Besides these simulations performed in non-relativistic regime, 
the transfer from tearing reconnection to a fully turbulent reconnection was indicated from the relativistic simulations of pulsar wind in Zrake (2016).
There it was found that the magnetic energy dissipation rate is 
insensitive to the grid resolution, showing the reconnection in the presence of turbulence is universal with respect to the unresolved physics. 
Based on the above studies, 
we expect a close similarity between the turbulence self-driven reconnection in non-relativistic and relativistic regimes.

\subsection{Modifications based on the ICMART model}

The pioneering model on GRBs built upon the turbulent reconnection is the ICMART model by ZY11. 
In this model, the interaction of magnetized slabs increases the degree of magnetic turbulence in the slabs, 
allowing magnetic fields to dissipate in a burst of turbulent reconnection.
 In other words, ZY11 invokes collision-induced magnetic reconnection and turbulence to interpret GRB prompt emission. Numerical simulations of collisions between magnetized blobs (Deng et al. 2015) revealed significant magnetic dissipation through reconnection. The magnetic dissipation efficiency can reach $\sim 35\%$, consistent with the analytical estimate of ZY11. The simulations also showed the existence of local Doppler-boosted regions due to reconnection, which is consistent with the mini-jets invoked in magnetic dissipation models of GRBs (Lyutikov \& Blandford 2003), which would shape the lightcurves of GRBs (Zhang \& Zhang 2014). The model entails a relatively large emission radius from the central engine, and has a list of features that match the observations very well (see \S \ref{sec:obs} below). 
Our present model is constructed based on the ICMART model and thus 
shares many common features with it that were described in detail in ZY11. 
Here we only focus on the modifications and discuss their necessity and significance.

As a major difference from the ICMART model, we introduce a more favorable mechanism of initiating turbulent magnetic 
reconnection. Namely, instead of collisions of magnetized slabs adopted in the ICMART model, we employ the kink instability, 
which naturally takes place in the relativistic and strongly magnetized jet of a GRB and inevitably induces turbulence and turbulent reconnection
(see \S \ref{ssec: kik}). 
The development of kink instability in a relativistic and Poynting-dominated jet is shown in numerical simulations
(Mizuno et al. 2012, 2014; ONeill et al. 2012).

Besides, as the theoretical core of both the ICMART model and our current model, 
turbulent reconnection has developed on more solid foundations. 
While ICMART model was suggested at the time when the theory of turbulent reconnection was supported only by non-relativistic simulations, 
by now new progress has been achieved in understanding relativistic turbulence.  
Compared to the original ZY11 publication, 
currently we have the numerical evidence that the turbulent reconnection is applicable to relativistic fluids (see e.g. Takamoto et al. 2015).

It is important to stress that the bursty feature of turbulent reconnection (LV99)
can account for the erratic behavior of GRB emission
(see \S \ref{sec:obs}). 
Consider a magnetically dominated low-$\beta$ plasma with weakly turbulent magnetic flux tubes coming into contact with each other. 
Initially, the magnetic reconnection proceeds at a slow pace (see Figure \ref{figure1}), 
as magnetic field lines are nearly laminar and the ratio of outflow region $\Delta$ to $L_x$ is very small.  
With the increase of $\Delta$, when the outflow Reynolds number becomes considerably larger than unity (see Eq. (\ref{Re})), 
the rising turbulence in the outflow will increase the fluctuations of surrounding magnetic field lines, inducing their higher level of 
wandering. 
This further extends the width of the outflow region $\Delta$ 
and increases the reconnection rate, as well as the energy injection in
the system. 
A higher level of energy injection and a higher $Re$ of the outflow both 
enhance the level of turbulence in the system. 
The above positive feedback can additionally enhance the level of turbulence that is initially excited by the kink instability, 
and leads to an explosion of reconnection. 
A quantitative model for such a process was presented 
for a non-relativistic low-$\beta$ reconnection in Lazarian \& Vishniac (2009).\footnote{Besides the application to GRB emission, 
this boot-strap turbulent reconnection
can also explain the formation of solar flares, as their existence requires both phases of slow and fast reconnection. 
In addition, the turbulence generated from the reconnection in one region can also trigger the reconnection in 
surrounding regions (LV99).
Such a process was reported in the observations of Sych et al. (2009, 2015) (see also Guti{\'e}rrez et al. 2017). }

\subsection{Comparison with other GRB models based on magnetic reconnection}

There exist several other GRB models that invoke magnetic reconnection as the origin of prompt emission. 
In the following we comment on how our model differs from those models.

Thompson (1994) envisaged a scenario of invoking mildly relativistic Alfven turbulence excited in the wind by reconnection, or by hydrodynamical instabilities triggered by magnetic tension. The reconnection process was discussed within the framework of the Petschek (1964) mechanism, which was later found unstable and not confirmed by numerical simulations. 
A photon spectrum is formed via Comptonization of thermal photons at a moderate or high scattering optical depth. The resulting spectrum is quasi-thermal, and the emission radius is close to the central engine. This is the earliest version of magnetic dissipative photosphere model in the GRB literature. Many authors further developed the magnetic dissipative photosphere model invoking magnetic reconnection below the photosphere (e.g. Drehkhahn \& Spruit 2002; Giannios 2006; Veres et al. 2013; Beniamini \& Giannios 2017). This model predicts a dominant photosphere emission component in the GRB prompt emission spectra, which may be consistent with some GRBs (e.g. GRB 090902B, Abdo et al. 2009b; Ryde et al. 2010; Pe'er et al. 2012), but may not explain those GRBs that do not show significant thermal emission component. 

Spruit et al. (2001) discussed a striped-wind magnetic field configuration with alternating polarity and argued that magnetic reconnection can happen continuously in the outflow both below and above the photosphere. They assumed that reconnection can proceed rapidly with local Alfven speed and argued that efficient $\gamma$-ray emission can be produced. The radiation spectrum was not calculated.

McKinney \& Uzdensky (2012) proposed a reconnection switch model of GRBs. They argued that as the GRB jet streams out, the comoving density in the jet decreases steadily. At a certain distance from the central engine, magnetic reconnection switches from the collisional regime (associated with Sweet-Parker reconnection) to the collisionless regime (associated with Petscheck reconnection) so that the reconnection speed increases rapidly. Significant magnetic dissipation occurs and a GRB is triggered. The switching distance could be below or above the photosphere radius, and the authors emphasized the possible enhancement of photosphere emission.

\section{Comparison with observations}\label{sec:obs}

The GRB prompt emission model outlined here shares many properties as the ICMART model, and has the advantage to interpret the observational data of at least some GRBs. In this section, we summarize how this model confronts many observational properties of GRBs:
\begin{itemize}
 \item Lightcurves: Observationally, GRB lightcurves are irregular and variable. Studies show that the lightcurves can be often decomposed into multiple ``pulses'' (Norris et al. 2005), each with durations of seconds. On the other hand, bursts can have rapid variability with a time scale as short as milliseconds. This ``fast'', spiky peaks often overlap with the ``slow'' pulse component (Gao et al. 2012). Similar to the ICMART model (ZY11), our kink-triggered GRB model interprets the slow pulses as individual kink-triggered events, while the fast spikes as due to comoving-frame mini-jets produced due to turbulent reconnection of individual units in a moderate-$\sigma$ jet. Monte Carlo simulations have shown that such a model can reproduce a variety of observed GRB lightcurves (Zhang \& Zhang 2014). 
 \item Spectra: Observationally GRB spectra have a dominant ``Band-function'' component (Band et al. 1993) with a typical low-energy spectral index $\alpha \sim -1$ (Preece et al. 2000; Nava et al. 2011; Zhang et al. 2011). Some bursts have a very hard spectral index ($\alpha > -2/3$), which is beyond the limit of the so-called synchrotron line-of-death (Preece et al. 200). In these cases, the spectra are likely of a thermal origin, which is consistent with emission from a fireball photosphere (M\'esz\'aros \& Rees 2000; Lazzati \& Begelman 2010). Observationally, the thermally dominated GRBs have been observed (Abdo et al. 2009b; Ryde et al. 2010; Pe'er et al. 2012), but for the majority of the GRBs, the thermal component is either sub-dominant (Guiriec et al. 2010; Axelsson et al. 2011) or not detectable (Abdo et al. 2009a; Zhang et al. 2016). This suggests that the GRB jets are Poynting-flux-dominated at the central engine, and likely in the emission region as well (Zhang \& Pe'er et al. 2009; Gao \& Zhang 2015). For these GRBs, the Band component is likely of a synchrotron radiation origin. Since photosphere emission is suppressed in these bursts, particles are likely accelerated in the turbulent reconnection region, rather than from internal shocks. At a large radius (beyond $10^{15}$ cm) from the central engine, magnetic field strength is low enough so that synchrotron cooling is no longer in the deep fast cooling regime. As the jet streams outwards, it is likely the comoving magnetic field strength continuously decreases with time. Fast cooling synchrotron spectrum in this model would deviate from the standard $\alpha = -3/2$ prediction, and give rise to a harder spectrum with $\alpha \sim -1$ (Uhm \& Zhang 2014; Geng et al. 2017b). Due to turbulent acceleration of electrons, the balance between cooling and acceleration of electrons would lead to a typical electron spectral index $p=1$, which gives rise to a photon power law spectral index $\alpha \sim -1$ (Xu \& Zhang 2017; Xu et al. 2017). Notice that these two ways to interpret $\alpha=-1$ make use of the two important predictions of our model: the large radius needed to have harder fast-cooling spectrum is consistent with requiring magnetized shells interacting to trigger kink instability, and turbulent acceleration needed to account for the $p=1$ is the natural consequence of turbulent reconnection induced from kink events. One interesting prediction of the model is that kink is easy to develop early on with the existence of the progenitor stellar envelope so that a bright thermal component may develop in the early phase of GRB. At later times, the jet would be Poynting flux dominated with emission powered by synchrotron radiation at a large emission radius. This is consistent with the recently observed bright, multi-episode GRB 160625B, which showed a transition from a fireball to Poynting flux dominated flow (Zhang et al. 2017; Troja et al. 2017).
 \item Spectral lag and $E_p$ evolution: The current picture invokes each kink event as one radiation unit. The observed broad pulse emission reflects the radiation history of the emission region as it streams outwards, rather than the history of the central engine activity. Such a picture naturally accounts for the observed spectral lag behavior (Norris et al. 2000) and $E_p$ evolution patterns (Lu et al. 2012), which is difficult to explain for the models invoking a small emission radius (Uhm \& Zhang 2016a). 
 \item Polarization: Polarized $\gamma$-ray emission has been claimed in some GRBs (Coburn \& Boggs 2003; Willis et al. 2005; Yonetoku et al. 2011, 2012). Even though with low significance, these observations nonetheless suggests that there is likely an ordered magnetic field component in the GRB emission region. This hypothesis is further supported by the detection of polarized optical emission shortly after $\gamma$-ray emission, either in the reverse shock region (Steele et al. 2009; Mundell et al. 2013) or in the internal prompt emission region (Troja et al. 2017). Our model can naturally account for all these observations.
 \item Neutrino upper limit: The IceCube Neutrino Observatory is placing progressively stringent upper limits on neutrino fluxes from GRBs (Aartsen et al. 2015, 2016, 2017), which greatly reduced the available parameter space of the models that invoke a small emission radius (e.g. the photosphere models and the internal shock models, Zhang \& Kumar 2013). Only the models that invoke a large enough emission radius. Since our kink-triggered magnetic dissipation model has the similar emission radius as that of ICMART, our model can comfortably satisfy the neutrino non-detection constraint.
\end{itemize}

\section{Discussion}

This paper presents the further step in the development of the ICMART model. 
The original model in ZY11 pioneered the concept of the turbulent magnetic reconnection for explaining major features of the GRB physics. The turbulent reconnection model that ZY11 appealed to was constructed in LV99 
and it was tested with non-relativistic 3D MHD simulations in Kowal et al. (2009). 
But at the moment of ZY11 publication, 
the properties of MHD turbulence in relativistic regime were mostly unclear, 
and the possibility of extending the LV99 model to the relativistic regime was also in question.
Nevertheless, ICMART (ZY11 for details) was able to successfully address a number of problems (e.g. low efficiency, electron fast cooling, electron number, weak or no photosphere emission in some GRBs, etc.) encountered by the internal shock model (Kumar 1999; Daigne \& Moshkovitch 1998; Ghisellini et al. 2000).

Since the publication of ZY11, the theoretical foundations of the GRB model based on turbulent reconnection 
in particular, the LV99 model, 
have been strengthened
(see Lazarian et al. 2016 for a review). 
This includes a better theoretical understanding of turbulent reconnection (see Eyink et al. 2012, Eyink 2015), 
more numerical testing (Kowal et al. 2012, Eyink et al. 2013, Oishi et al 2015, Beresnyak 2017, Kowal et al. 2017), and 
more observational evidence (e.g. Lalescu et al. 2015). 
Most importantly, the theory of relativistic MHD turbulence has been advanced 
(see Takamoto \& Lazarian 2016, 2017), and the relativistic turbulent reconnection has been demonstrated numerically in TIL15. 
These updates make it important to revisit the ICMART model.

On the other hand, from the observational front, many new observations since ZY11 
support the general picture of the ICMART model at least in some (probably in most) GRBs. 
These include the polarized $\gamma$-ray and optical emission of GRB prompt emission and early afterglow (Yonetoku et al. 2011, 2012; Mundell et al. 2013; Troja et al. 2017), the progressively tight upper limits of the neutrino flux from GRBs (Aartsen et al. 2015, 2016, 2017), as well as evidence of bulk acceleration and/or anisotropy in the GRB emission region (Uhm \& Zhang 2016a,b; Geng et al. 2017a). This motivated us to further develop the ICMART model in terms of more robust reconnection physics and an alternative (and probably more realistic) triggering mechanism.

The present paper addresses the above observational challenges by presenting and quantifying a new mechanism of triggering 
flares of reconnection. 
It appeals to the kink instability, which 
alters the original configuration to that prone to magnetic reconnection. 
This significantly improves the ability of the model to explain observational data. 
A bursty emission model due to turbulent reconnection is discussed in detail in view of the latest developments in reconnection physics. 
This lays a solid ground to the sketchy picture delineated in the ICMART model of ZY11. 
\\
\\

AL acknowledges the support the NSF grant DMS 1622353 and AST 1715754 and thanks Misha Medvedev for helpful discussions.
BZ acknowledges NASA NNX15AK85G for support.
SX acknowledges the support for Program number HST-HF2-51400.001-A provided by NASA through a grant from the Space Telescope Science Institute, which is operated by the Association of Universities for Research in Astronomy, Incorporated, under NASA contract NAS5-26555.

\end{document}